\documentclass[article]{revtex4-2}  

\usepackage{hyperref}
\usepackage{natbib}
\usepackage{etoolbox}
\apptocmd{\thebibliography}{\raggedright}{}{} 

\usepackage[utf8]{inputenc}
\usepackage{subfiles}
\usepackage{amsfonts}
\usepackage{graphicx}
\usepackage{physics}
\usepackage{float}
\usepackage{amsmath}
\usepackage{mathtools}
\usepackage[english]{babel}
\usepackage{csquotes}

\newcommand{\opa}{a}
\newcommand{\opada}{a^\dagger}

\newcommand{\mean}[1]{\langle #1 \rangle}
\newcommand{\beq}{\begin{equation}}
\newcommand{\eeq}{\end{equation}} 
\newcommand{\bea}{\begin{eqnarray}}
\newcommand{\eea}{\end{eqnarray}}

\usepackage{xcolor}

\newcommand{\uba}{Universidad de Buenos Aires, Facultad de Ciencias Exactas y Naturales, Departamento de Física. Buenos Aires 1428, Argentina}
\newcommand{\ifiba}{CONICET - Universidad de Buenos Aires, Instituto de Física de Buenos Aires (IFIBA). Buenos Aires 1428, Argentina}

\begin{document}
\title{Defining coherent states: why must they be eigenstates of the annihilation operator?}

    \author{Juan Pablo Paz}
    \affiliation{\uba}
    \affiliation{\ifiba}

    \author{Augusto J. Roncaglia}
    \affiliation{\uba}
    \affiliation{\ifiba}

    \begin{abstract}
       This is a pedagogical paper where we present a physically motivated approach to introduce the coherent states of a harmonic oscillator from which it is simple to rigorously derive their mathematical definition.  We do this in two different ways that turn out to be equivalent  but emphasize two related but different aspects of classicality.  First, we  analyze which are the quantum states that are the closest one can get to a point in phase space and demonstrate the validity of the following theorem: $(i)$ The product of the uncertainty in position and that of momentum saturates the bound imposed by Heisenberg uncertainty relations  for all times if and only if the state is an eigenstate of the annihilation operator. 
       Second, we analyze the way in which the difference between the expectation value of the energy and the energy associated with the expectation values of position and momentum depends on the state, and show the validity of the following theorem $(ii)$ the difference between the expectation value of the energy and the energy associated with the expectation values is minimal 
       if and only if the state is an eigenstate of the annihilation operator. We also show that the reason why coherent states are chosen as the most classical ones by the decoherence process induced by coupling the particle to an environment in the standard Quantum Brownian motion model, is precisely due to the validity of the two above theorems.
\end{abstract}

\maketitle
\section{Introduction} 
\label{sec:introduction}
In this paper we will present and discuss  a strategy to define the coherent states of the harmonic oscillator~\cite{schrodinger1926stetige,glauber1963coherent} starting from a physically motivated principle from which one can derive the rigorous mathematical definition. This is, in our opinion, the best way to introduce this important class of states and  motivate their study. 
There is no need to argue much about the importance of coherent states, which are the most classical states of the harmonic oscillator. In fact, they are widely recognized as fundamental to understand the behavior of many natural systems. As an example, it is enough to mention that they are used to describe the state of the light emitted by a laser source, as they are so important, they appear in every text book on quantum mechanics~\cite{Cohen-Tannoudji:1977,merzbacher1968,sakurai1985quantum,GalindoQuantumMechanics,ballentine,Weinberg2015,gottfried_yan_quantum_mechanics}.
Coherent states have a very simple mathematical definition: they are eigenstates of the annihilation operator $\opa$ (see below for definitions), they are usually denoted as $\ket \alpha$ and satisfy the equation  $\opa \ket\alpha = \alpha \ket\alpha$, for a certain complex number $\alpha$ that fully characterizes each state. But the question we discuss in this paper is: why should we study the eigenstates of the annihilation operators to begin with. Or, in other words, we ask if there is a physically motivated guiding principle, that in one way or another can be used to characterize classical behavior, from which one can  derive the mathematical formulation of coherent states.  In fact, we look for an argument stated in the following way: ``if the physical condition X is satisfied, then the state must be an eigenstate of the annihilation operator'' (for a somewhat related approach, independent of ours, see~\cite{martin2021searching}). In what follows, we will present two formulations of this type, which are fully based on a clear physical requirement and enable us without any “hand waving” to conclude that the eigenstates of annihilation operator are the states we should carefully look at.

The paper is organized as follows: In Sec.~\ref{sec:definitions}, we briefly introduce our notation and review the essential features of the harmonic oscillator that are typically covered in text books. In that section we skip all demonstrations and basically state the most important results we need to  introduce the coherent states. All these demonstrations (including, for example, the calculation of eigenvalues and eigenstates of the Hamiltonian) are covered with great detail and quality in standard references. In Sec.~\ref{sec:Review}, 
we review the way in which coherent states are introduced in text books without trying to accomplish the impossible goal of going through all of them, but focusing only the most influential ones according to our point of view. 
In Sec.~\ref{sec:coherent_states} we present the two arguments we devised in order to satisfy the above stated requirement. As we mentioned, they are equivalent being one of them based on imposing the state to be maximally localized in phase space for all times while, the other one is based on the idea of classical orbits being followed in the closest possible manner. In Sec.~\ref{sec:deco} we present a short discussion on the connection between our criteria and the one based on selecting coherent states as being the least affected by environmental interactions in the Quantum Brownian motion model.
In Sec.~\ref{sec:conclusions} we present a summary of the results and conclusions.

\section{Definitions and notation}
\label{sec:definitions}
So let us first review some of the notation we will use. We will consider a harmonic oscillator with a Hamiltonian
\beq
H=\frac{p^2}{2m}+\frac{m\omega^2}{2}x^2.
\eeq
This is a mechanical oscillator with mass $m$ and frequency $\omega$ but other types of oscillators (like the ones associated with modes of the electromagnetic field), can be studied in the same way. As it is customary, the annihilation operator $\opa$ is defined as
\beq
\opa = \frac{1}{\sqrt 2}\left(\frac{x}{\sigma}+i\frac{\sigma}{\hbar} p\right)
\eeq
Where $\sigma=\sqrt{\frac{\hbar}{m\omega}}$ is the length scale one can define with Planck constant and the oscillator parameters.
In terms of $\opa$ and $\opada$ (which obey the canonical commutation relation $[\opa,\opada]=1$) the Hamiltonian reads as
\beq
H=\hbar \omega \left(\opada  \opa + \frac{1}{2}\right).
\label{eq:Hamiltonian}
\eeq

The typical presentation starts by defining the number operator $N= \opada \opa$  in terms of which the Hamiltonian is simply $H= \hbar \omega (N+1/2)$ and to use the commutation relations
$[N,\opa]=-\opa$, $[N,\opada]=\opada$,
to show that $\opa$ and $\opada$ enables one to jump one step down or up in the energy spectrum of $H$. From this observation, the demonstration that the eigenvalues of $N$ are positive integers (which we denote by $n$) and that the energy eigenstates $\ket{\phi_n}$ can be built from the ground state by  acting $n$-times with $\opada$ follow almost directly.
Moreover, typical text books also show how to compute expectation values of products of powers of position and momentum operators by writing them in terms of $\opa$ and $\opada$ and using the simple matrix elements these operators have in the energy eigenstates basis.

For our purpose, it will be useful to remember here how to write some expectation values in terms of the expectation value of creation and annihilation operators. From its very definition shown above we have
\beq
\mean{a}= \frac{\mean{x}}{\sqrt{2}\sigma}+i  \frac{\mean{p}\sigma}{\sqrt{2}\hbar}= \alpha
\eeq
therefore the energy associated with the expectation values of $x$ and $p$ is also easily obtained from $\alpha$ as:
\beq
\bar E \equiv \frac{\mean{p}^2}{2m}+\frac{m \omega^2}{2}  \mean{x}^2=\hbar \omega \abs{\alpha}^2
\label{eq:Ebar}
\eeq

Finally, it is convenient to mention a few simple aspects of the temporal evolution of a harmonic oscillator that are usually find in all text books. 
The one we will find more useful for our purpose is the observation that the Heisenberg equations for the operators $x$ and $p$ are identical to Newton's equations, i.e
\bea
i \hbar \dot x&=& [x,H]= i \hbar \frac{p}{m}\nonumber \\
i \hbar \dot p &=& [p,H]=- i \hbar m\omega^2 x
\eea
and therefore they can be explicitly solved. From now on, we will denote the Heisenberg operator at time $t$ as $x_t$ to make our notation more compact. Thus, the Heisenberg operators are simply written in terms of Schr\"odinger operators $x_0$ and $p_0$ as:
\bea
x_t&=& x_0 \cos(\omega t)+\frac{p_0}{m\omega} \sin(\omega t)\nonumber \\
p_t&=& p_0 \cos(\omega t)-{m\omega x_0}\sin(\omega t)
\eea
Using these equations, it is simple to compute the variance of position and momentum as a function of time. In fact, we can write them in terms of the initial values of the dispersions $\Delta x_0 \equiv \sqrt{\mean{x_0^2}-\mean{x_0}^2}$ and $\Delta p_0$, and they involve also a dependence on the initial position momentum correlation function $K(x_0,p_0)=\mean{x_0p_0+p_0x_0}/2- \mean{x_0} \mean{p_0}$. They read as
\bea
(\Delta x_t)^2&=& (\Delta x_0)^2 \cos^2(\omega t) + \frac{(\Delta p_0)^2}{m^2\omega^2} \sin^2(\omega t) + \frac{K(x_0,p_0)}{m\omega} \sin(2 \omega t) \nonumber \\ 
(\Delta p_t)^2 &=& (\Delta p_0)^2 \cos^2(\omega t)+ m^2\omega^2  (\Delta x_0)^2 \sin^2(\omega t) - m\omega K(x_0,p_0) \sin(2 \omega t)
\label{eq:dispersions}
\eea

\section{Review of the Literature}
\label{sec:Review}

Before reviewing what  text books say about coherent states, it is worth mentioning some fundamental papers of great historical value in the development of this concept.
Coherent states were introduced in 1926 by E. Sch\"ordinger~\cite{schrodinger1926stetige} who denoted them as semiclassical states, he argued these state should satisfy the following properties: the expectation values of position and momentum should follow classical trajectories, the wave packet should not spread (and therefore the dispersions $\Delta x$ and $\Delta p$ should remain constant) and, finally, the expectation values of the energy should stay close to the classical one (which is naturally constant). The first of these conditions is trivially satisfied by all quantum states of the harmonic oscillator, as Heisenberg equations are identical to the classical ones. Thus, the coherent states of Sch\"ordinger are really defined only by the two last properties. However, as Sch\"ordingers' paper was written in the early  days of quantum mechanics the relation between these semiclassical states and the eigenstates of the annihilation operator was not mentioned.  The relevance of coherent states for the harmonic oscillator became more noticeable after the work of R. Glauber~\cite{glauber1963coherent} who defined the coherent states as the eigenstates of the positive frequency part of the quantum field which, together with the fact that the state is a product of the ones associated with the different modes, imply the definition of coherent states as eigenstates of all the annihilation operators.

Let us now briefly review how modern textbooks approach the definition of coherent states.
Most textbooks introduce coherent states as the most classical quantum states of the harmonic oscillator.
However, at the time of defining them, two approaches are generically followed: the first one is defining them by using arguments based on some kind of semiclassical limit.
The second one, is defining them as the eigenstates of the annihilation operator and then obtaining all the desired properties. 
The celebrated and brilliant book of C. Cohen Tannnoudji and co-workers~\cite{Cohen-Tannoudji:1977}  is very close to what we aim for but does not fully accomplish our goal. In fact, in the corresponding section devoted to the analysis of “semiclassical coherent states” the authors argue that when the energy is large, the ground state energy can be neglected and  the quantum Hamiltonian can be approximately written as $H\approx \hbar \omega \mean{\opada \opa}$. This can be  identified with the energy associated to the expectation values of position and momenta as $H\approx\hbar \omega \abs{\alpha}^2$. 
Then, the identity $\hbar \omega \mean{\opada \opa} =\hbar \omega |\alpha|^2 $, which is mathematically  equivalent to $\mean{\opada \opa}-\abs{\alpha}^2=0$ implies that the state must be an eigenstate of the annihilation operator. As we will show below, the same argument holds without any hypothesis related with the semiclassical limit, coherent states are the ones that minimize the difference between the expectation value of the energy and the energy associated with the expectation values of position and momentum.

Other very valuable books approach the  definition of coherent states from a more mathematical perspective simply stating that they are defined as the eigenstates of the annihilation operator and they show from this that they are, in fact, the closest one can get to a classical state respecting the principles of quantum mechanics. This is the approach presented by J.J. Sakurai \cite{sakurai1985quantum},  A. Galindo and P. Pascual~\cite{GalindoQuantumMechanics}, E. Merzbacher~\cite{merzbacher1968}, or S. Weinberg~\cite{Weinberg2015} for example. While for instance L. Ballentine~\cite{ballentine} or  K. Gottfried and T-M. Yan~\cite{gottfried_yan_quantum_mechanics}  define them as displacements of the ground state of the harmonic oscillator.  Even books specialized on coherent states and their applications such as the notable one by A. Perelomov~\cite{Perelomov1986} introduce them in this way (or in some more abstract manners generalizing them to other groups different from the Heisenberg Weyl one). Books on quantum optics such as the well known one by D. Walls and G. Milburn~\cite {walls2008quantum} or that of P. Knight~\cite{GerryKnight2004} also do it in this manner.

\section{Introducing coherent states using a rigorous physical definition}
\label{sec:coherent_states}
Here, we will  present the most important result of this paper: a definition of  coherent  states based on simple but rigorous physical principles. As we mentioned, we will  accomplish this goal
in two apparently different ways that turn out to be equivalent. Both principles that we discuss here are based on the idea that coherent states should capture the notion of classicality. Thus, this is what we are looking for, the set of states of a harmonic oscillator that behave in the most classical way. As we also mentioned, as all quantum states are such that their expectation values of position and momentum obey Newtons equations, the behavior of these expectation values cannot be directly and naively used to define a preferred family of states.

First, we will analyze one of the conditions that is a characteristic feature of classicality: the state should describe something that looks as close as possible like a point in phase space.  For this reason, we will analyze the states that saturate Heisenberg inequality at all times. That is, we will find out which are the quantum states that behave in such a way that the equation
\beq
\Delta x_t \Delta p_t = \frac{\hbar}{2}\quad  \forall t
\label{eq:Heisenberg}
\eeq
Our analysts is, as shown below, very simple from the mathematical point of view but leads us to very important conclusions that can be captured in the following Theorem:

\emph{Theorem 1}: $\Delta x_t \Delta p_t= \hbar/2$  $\ \forall t$ $\iff$ the state $\ket\psi$   is such that $\opa \ket \psi= \alpha \ket \psi$ where the eigenvalue $\alpha$ is a complex number.

 The fact that the right hand side  of the above enunciated theorem implies the validity of the left hand side is quite  simple and  can be found in all text books. Thus, one can prove that the eigenstates of the annihilation operator saturate the bound imposed by Heisenberg inequality for all times. The non-trivial part is the direct implication (that is to say, the proof of the fact that if the Heisenberg inequality is saturated for all times, then the state must be an eigenstate of the anihillation operator). To prove this we proceed as follows: For convenience we will squared the Eq.~\eqref{eq:Heisenberg} and rewrite it in the following equivalent way: 
$4 \left(\frac{(\Delta p_t)^2}{2m}\right) 
 \left( \frac{m\omega^2 (\Delta x_t)^2}{2}\right)=\left(\frac{\hbar\omega}{2}\right)^2$. When written in this manner the equation suggests the introduction of the kinetic energy operator, $T=\frac{p^2}{2m}$ and the potential operator $V=\frac{m\omega^2}{2}x^2$. Thus, the term $\Delta p_t^2/2m$ is just the difference between the mean value of the kinetic energy ($\mean{T}_t = \mean{p^2_t}/2m$) and the kinetic energy of the expectation value ($\bar T _t= \mean{p_t}^2/2m$). That is to say $\delta T =\mean{T}-\bar T =(\Delta p)^2/2m$. Similarly, the term $m\omega^2\Delta x^2/2$ can be associated with the difference between the mean value of the potential energy ($\mean{V}$) and the potential energy of the mean values ($\bar V=m\omega^2\mean{x}^2/2$), i.e. $\delta V = \mean{V}-\bar V=m\omega^2\Delta x^2/2$.
Thus, the Heisenberg identity Eq.~\eqref{eq:Heisenberg} can be expressed as:
\beq
4\, \delta T_t\, \delta V_t = \left(\frac{\hbar \omega}{2}\right)^2
\eeq
 Using the algebraic relation $(b +c)^2-(b-c)^2=4bc$,  for $b=\delta T_t$ and $c= \delta V_t$, the equation can be rewritten as: 
 \beq
 \left(\delta H_t\right)^2-\left(\delta L_t\right)^2=
\left(\frac{\hbar \omega}{2}\right)^2
\eeq
 where $H=T+V$ is the Hamiltonian and $L=T-V$ is the Lagrangian. It is clear that $\delta H_t\equiv\delta H=\mean{H} -\bar E$ does not depend on time as it is the difference between two time independent quantities:  $\mean{H}$ which is trivially constant, and the energy of the expectation values $\bar E$ is conserved by Newton's equations. Instead $\delta L_t$ depends on time and its expression can be obtained from Eqs.~\eqref{eq:dispersions}:
 \beq
 \delta L_t = \delta L_0 \cos(2\omega t)+\frac{\omega K(x_0,p_0)}{2}\sin(2\omega t)
 \eeq
 Therefore for the Heisenberg inequality to be saturated at all times $\delta L_t$ must vanish, which requires that  $\delta L_0=0$ and $K(x_0,p_0)=0$. The first identity simply implies that ${\Delta p_0=m\omega \Delta x_0}$, and the second reflects the absence of initial correlations between position and momentum. When these conditions are satisfied, the saturation of the Heisenberg inequality implies that:
\beq
\delta H=\frac{\hbar \omega}{2}
\eeq
In turn, $\delta H $ can be easily written in term of the creation and annihilation operators (see Eqs.~\eqref{eq:Hamiltonian} and \eqref{eq:Ebar}).  Doing this, we can write 
\bea
\delta H &=& \hbar \omega \left(\mean{\opada \opa} +\frac{1}{2}- \abs{\alpha}^2\right) \nonumber\\
&=&\hbar \omega \left(\|(\opa- \alpha)\ket\psi\|^2 +\frac{1}{2}\right)
\label{eq:deltaHmin}
\eea
where $\|\ket\phi\|=\sqrt{\braket{\phi}{\phi}}$ is the usual norm of the vector $\ket\phi$.
As  Heisenberg inequality saturates at all times if $\delta H=\hbar \omega/2$, we obtain that the states $\ket \psi$ inducing this saturation are such that $\|(\opa- \alpha)\ket\psi\|^2=0$ which in turn implies that
\beq
\opa \ket \psi= \alpha \ket \psi.
\eeq 
This completes the proof of the Theorem (the reverse implication is left to the reader as an exercise that can also be found in all text books).

As we mentioned above there is another way to define coherent states that turns out to be equivalent to the previous one. In fact, if we simple look for the states that minimize $\delta H$, i.e. the difference between the expectation value of the energy and the energy of the expectation values we immediately find that such states should be eigenstates of the annihilation operator. In fact, this is the statement of our second theorem. 

\emph{Theorem 2:} $\delta H$
is minimal $\iff$ $\opa \ket \psi= \alpha \ket \psi$.

Again the reverse implication is trivial and now in view of the above demonstration of Theorem 1 it can be proven in a very simple way. Thus, as according to Eq.~\eqref{eq:deltaHmin}, we can write $\delta H$ as
\beq
\delta H =\hbar \omega \|(\opa- \alpha)\ket\psi\|^2 +\frac{\hbar \omega}{2}
\eeq
Therefore, the minimal value of $\delta H$ is attained when the norm in the previous equation vanishes. Which implies that such states should be eigenstates of the annihilation operator.

It is interesting to notice that in the above demonstrations  the Hamiltonian and the Lagrangian, which  are both key characters of classical mechanics, appear but their role is rather different. In fact, our argument was based on writing the product of the position and momentum dispersions as:
\beq 
\left(\Delta x_t \Delta p_t\right)^2=\left(\frac{\delta H}{\omega}\right)^2-\left(\frac{\delta L_t}{\omega}\right)^2,
\eeq
From this, $\delta L_t=0$ is a necessary but not a sufficient condition to saturate Heisenberg inequality at all time. Thus, the expectation value of the Lagrangian is equal to the Lagrangian of the expectation value, the product of the dispersions is constant but not necessarily minimal. An example of states satisfying this condition is provided by the energy eigenstates $\ket{\phi_n}$, for which $\Delta x \Delta p = \hbar \omega (n+1/2)$.
Instead, the condition that $\delta  H$ attains its minimal implies that  $\delta L_t=0$ and that Heisenberg inequality saturates at all times. This is interesting and makes clear that one can rigorously define coherent states by using the minimization of the Hamiltonian difference $\delta H$ but not the Lagrangian difference $\delta L$ (nor the action difference, which can be obtained by integrating $\delta L$ in time). Lagrangian and Hamiltonian are two of the leading characters of classical mechanics but in order to define the classical states in quantum mechanics they clearly play a rather different role. 

\section{Coherent states, Decoherence and Predictability}
\label{sec:deco}

The emergence of coherent states as a “special set of states” has been approached in the past from a different perspective that,   however, has  many points in common with the one we described in the previous Section. Thus, the existence of a preferred set  of states in the Hilbert space of a physical system is the problem one faces when trying to explain the emergence of classicality. In particular, 
for a measurement apparatus to behave classically it should never be found in a quantum state that is a superposition of the so-called “pointer states”~\cite{zurek1981pointer}. In recent decades it became clear that there is a physical process that explains why this happens in the macroscopic world.  The existence of this process, which is known as “decoherence”~\cite{zurek1991decoherence,paz1999environment,zurek2003decoherence},  is not related with the size of the system but to its openness.  For systems that interact with  many other degrees of freedom that constitute an environment in which the system is immersed in, decoherence is the most effective process that takes place  in a  typically short timescale.   

The explanation that decoherence gives to the emergence of a preferred set of states is the following: this set is dynamically selected by the combined effect of the interaction Hamiltonian between the system and the environment together with the Hamiltonian that characterizes the evolution of the system and the environment separately. As a result, in typical situations, a set of states   is selected in which the system can exist in a relatively stable way. These states are the ones that are minimally affected by the interaction with the environment, while superpositions of them rapidly decay into mixtures. As a result, quantum coherence in the system is lost and the purity of the states diminishes. Decoherence, therefore, explains the transition from a state where the system (an apparatus, for example) starts in a state in which is “here and there” at the same time into a state where it is “here or there” with some probabilities. 

The physics of the process of decoherence has been extensively studied in recent decades  not only because it is a key ingredient to explain the emergence of a classical world out of a fundamentally quantum substrate. More recently, in fact, it became clear that decoherence is the enemy to defeat in order to achieve quantum information processing.
In a paper~\cite{zurek1993coherent} authored by W. Zurek and one of us (JPP)  a criterion for finding the preferred set in a systematic manner was used for the first time. Such criterion, that was proposed earlier in \cite{zurek1993preferred}  was denoted as the “predictability sieve”.
The criterion states that one should look at the states which are the most stable against the environmental effect, the ones that loose the least amount of purity and, in a precise sense, become less entangled with the environment. 
In order to apply it operationally one would have to find out which are the states of the system that loose the smallest amount of purity (for example) due to the interaction with the environment. A preferred set emerges if the application of this criterion gives a stable answer for a range of times which is dynamically interesting (of course, one could find that a different set of states is minimally affected by environmental interaction for every time $t$ but this is not what we want, a preferred set emerges only if the same answer is found for a relevant range of times).
Doing this in practice is not at all simple since the full dynamics of the combined universe formed by the system and its environment is hard to solve except in a few exceptional cases. One of them is the so-called Quantum Brownian Motion (QBM) model where exact solutions are available~\cite{hu1992quantum}. In that case, the  system consists of a single particle that moves in one dimension while interacting with an environment formed by a collection of an infinite number of harmonic oscillators. The system environment interaction is assumed to be bilinear in the coordinates of the system and those of the environment.  The predictability sieve could be implemented after some simplifying approximations (see below) and a notable answer emerged: Coherent states are selected by decoherence (in fact, the title of the above mentioned paper is precisely “Coherent states via decoherence”).

In the context of the discussion we presented in the previous Sections, the predictability sieve is a valid physical criterion that could be used to define coherent states. But why is it that this apparently different type of principle, that involves the interaction between the system and an environment, yield the same answer we found above. Here, we will give a simple explanation based on the simplest model one could use to solve the evolution of the QBM model. In fact, the quantum state of the Brownian particle (the system in this case) must be described by a density matrix, that is denoted as $\rho$ (a positive operator that has unit trace). This state does not satisfy Schr\"odinger's equation precisely because the interaction with the environment induces a non-unitary evolution that does not preserve, for example, the purity of the state. Which can be measured as $\xi=\Tr[\rho^2]$, notice that purity is unity for pure states (where $\rho^2= \rho$) and decays for mixed states. So, the simplest implementation of the predictability sieve is to look for the initial states of the system that minimize the decay of $\xi$. To do this, we need an evolution equation for the density matrix of the system, which is usually known as a “quantum master equation”~\cite{hu1992quantum}. There is a vast literature on such equation but it is not the goal of this paper to review any of this. Instead, we will simply state the equation in a remarkably simple and powerful physical limit: the so-called “low damping and high temperature limit” of the QBM model with an environment characterized by an ohmic spectral density (that induce, for the evolution equation of the expectation values of position and momentum, a friction force proportional to the velocity of the particle) and an infinite high frequency cutoff. In such case, the equation reads as
\beq
\dot\rho=-\frac{i}{\hbar}[H,\rho] -\frac{i\gamma}{\hbar} [ x,\{p,\rho\}]- \frac{2m \gamma  k_B T}{\hbar^2} [x,[x,\rho]].
\eeq
The first term in the right hand side refers to the unitary part of the evolution while the second and the third ones are respectively associated with the two most important effects the interaction with the environment produces: dissipation (energy loss) and diffusion (purity loss or entropy production). Before proceeding, it is convenient to mention that the above equation is just an approximation to the exact master equation for the QBM model and, as such, can only be applied in certain circumstances. It does not have the so-called Lindblad form (that is the type of master equation  that guarantees the preservation of positivity and trace) but this is not a problem unless we try to apply it beyond its range of validity. For example, the above equation was derived in the high temperature limit which is valid when $k_BT$ is the largest energy scale available in the problem (for example, it should be satisfied that the following inequalities hold: $k_BT \gg \hbar \omega$, $\hbar \gamma$ and $k_BT\gg \hbar^2/2m \Delta x^2$). In any case, the master equation enables us to obtain  a simple expression for the evolution of the purity of the system. Thus, using that $\dot \xi= 2 \Tr[\rho \dot \rho]$ it is simple to show that
\beq
\dot \xi=  2\gamma \xi - 8 \gamma \Tr(\rho^2 x^2 - \rho x \rho x)/\lambda_T^2,
\eeq
Where the thermal de Broglie wavelengths is defined as $\lambda_T=\hbar/\sqrt{m k_BT}$. To continue we use the following approximations: we assume the damping rate $\gamma$ is very small, but the product $\gamma T$ is finite (this defines the weak coupling, high temperature limit) and we assume in the right hand side of the above expression that the state is approximately pure replacing $\rho= \ketbra{\Psi}{\Psi}$. In such case we get
$\dot \xi= - 8 \gamma \Delta^2 x/\lambda_T^2$,
where $\Delta x$, as usual is the dispersion in position. Using the expression for the position operator as a function of time given above and  averaging the right hand side over a period of the oscillation, we obtain an estimation for the rate at which purity decays that reads as
\beq
\dot \xi= - \frac{8\gamma k_BT}{(\hbar \omega)^2} \left(\frac{(\Delta p)^2}{2m}+ \frac{m\omega^2}{2} (\Delta x)^2\right),
\eeq
That, in view of the discussion presented in the previous Section, can be simply written in terms of $\delta H$, the difference between the expectation value of the energy and the energy of the expectation values. In fact, we can write
\beq
\dot \xi= -\frac{8\gamma k_BT}{\hbar \omega} \frac{\delta H}{\hbar \omega}.
\eeq
Therefore, the states that minimize the purity decay are precisely the ones that minimize the difference $\delta H$, which are (as shown above) the eigenstates of the annihilation operator. For those states, the decay of purity is
$\dot \xi= -4\gamma (k_BT/\hbar \omega)$.  This shows that, as stated in the title of the above mentioned paper, coherent states (eigenstates of the annihilation operator) are selected via decoherence.

\section{ Concluding remarks}
\label{sec:conclusions}

In the paper, we presented two different (but closely related) physical criteria that can be used to define the so-called coherent states. Their virtue is that the physical insight they provide (phase space localization or maximal closeness to the classical trajectories followed by expectation values of position and momentum) rigorously imply that the coherent states are eigenstates of the annihilation operator. As we mentioned in the introduction, we looked for an argument of the following form: ``if the physical condition X is satisfied then the state must be an eigenstate of the annihilation operator''. Our work can be  seen as a way to present two versions of the above mentioned ``physical condition X'' that characterized classicality. Therefore, we suggest to define coherent states by one of the following statements: ``coherent states are the ones that minimize the product $\Delta x\Delta p$ for all times", or ``coherent states are the ones that minimizes the different between the expectation value of the energy and the energy of the expectation values". As our previous considerations showed, once we define the coherent states in one of these two physically motivated manners, we can say that they both imply that such states are eigenstates of the annihilation operator and that these are the states we should analyze with great mathematical rigor, as most books do.

Finally, in Sec.~\ref{sec:deco} we presented an argument to show that open quantum  systems interacting with realistic environments (which in this case is the quantum Brownian motion model, that describes a particle coupled with an ohmic environment in the limit of very low damping and very high temperature) may be used to explain why is it that coherent states are the ones where the particle is typically found. All the other states decay rapidly into mixtures of coherent states, which are the most robust against the interaction with the environment. We should point out that this conclusion significantly depends on the properties of the model (i.e. on the environmental spectral density, on the relation between the system and the environmental dynamical timescales, on the strength of the interaction, etc). In fact, a variety of situations have been analyzed where, depending on the case, decoherence can select (through the predictability sieve criterion) either coherent states or eigenstates of the interaction Hamiltonian~\cite{zurek1981pointer} (as it is the case in the very short time limit of the model described in Sec.~\ref{sec:deco} where the purity loss seems to be minimized by position eigenstates for which $\Delta x=0$) or by the eigenstates of the systems' Hamiltonian~\cite{paz1999quantum} (a situation that is typically achieved for spin systems when the environmental frequencies are typically smaller than the ones of the system). The model we discussed, the QBM model, is a realistic description for the evolution of a particle immersed in an environment which induces a friction force proportional to the velocity and when the dominant effect is not the energy loss but the loss of quantum coherence. in such case, it is clear that the coherent states, as defined in this paper as the ones minimizing  the product of position and momentum dispersion for all times (or, equivalently, the energy difference $\delta E$) are the ones selected via decoherence.

\section*{Acknowledgments}
JPP would like to acknowledge support from the ICTP through the Associates Programme and the Simons foundation through the grant 284558FY19.
AJR would like to acknowledge support from the ICTP through the Associates Programme (2022-2027).



%

\end{document}